\documentclass[runningheads]{llncs}
\usepackage{graphicx}
\usepackage{algorithmic}
\usepackage{algorithm}
\usepackage{xcolor}
\usepackage{enumitem}
\usepackage{graphics}
\usepackage{subfigure}
\usepackage{wrapfig, epsfig}
\usepackage{makecell}
\usepackage{lipsum}
\usepackage{mwe}

\usepackage{amsmath}

\begin{document}
\title{Top-k Dynamic Service Composition in Skyway Networks\vspace{-0.4 cm}}
%
%
%
\author{Babar Shahzaad \and
Athman Bouguettaya}
\authorrunning{B. Shahzaad et al.}
\institute{The University of Sydney, Sydney NSW 2000, Australia\\
\email{\{babar.shahzaad, athman.bouguettaya\}@sydney.edu.au}}

\maketitle              
\vspace{-0.6 cm}
\begin{abstract}
We propose a novel top-k service composition framework for drone services under a dynamic environment. We develop a system model for formal modelling of drone services in a skyway network. The composition process is accomplished in two phases, i.e., computing top-k compositions and extending and ranking top-k compositions using probabilistic wait and recharge times under congestion conditions. We propose a top-k composition algorithm to compute the best service composition plan meeting user's requirements. A set of experiments with a real dataset is conducted to demonstrate the effectiveness of the proposed approach.
\vspace{-0.2 cm}
\vspace{-0.1 cm}
\end{abstract}

\section{Introduction}

Drones have gained significant attention in recent years due to their potential benefits for a multitude of civilian applications \cite{8682048}. The use of drones will play a paramount role in enabling new services in various domains such as disaster management, remote sensing, and delivery of goods \cite{2021335}. Drones provide safe, contactless, and more resilient alternatives to deliver goods in remote locations \cite{9086010}. Many start-up companies such as FlyTrex and large companies such as Amazon and Google are investing in the use of drones for delivery services \cite{Aurambout2019}.

The \textit{service paradigm} \cite{Bouguettaya:2017:SCM:3069398.2983528} offers a powerful mechanism to abstract the capabilities of a drone {\em as drone services}. As any other service, a drone service is defined by its \textit{functional} and \textit{non-functional} properties \cite{alkouz2020formation}. In this instance, the functional property represents the {\em transport of a package} from a node (e.g., warehouse rooftop) to another node (e.g., customer's building rooftop) in a {\em skyway} network. The non-functional (i.e., \textit{Quality of Service} (QoS)) properties of a drone service represent such attributes as the payload capacity, flight range, battery capacity, etc. A {\em skyway network} is defined as a set of connected nodes representing take-off and landing stations \cite{10.1145/3460418.3479289}. Each node may concurrently act as a recharging station. The transport/delivery of a package by a drone along a line segment that directly connects two nodes represents an atomic service abstraction. An instantiation of this service abstraction is the transport of a package by a specific drone between two named nodes that are connected by a direct segment, operating under a set of requirements/constraints.

A single drone service may not guarantee the direct delivery of a package from a warehouse to a customer's desired location due to flight range limitations, flight regulations, battery life, etc. Therefore, drone service {\em composition} is required to ensure successful package delivery. An optimal drone service composition is defined as the selection of the best drone services in a skyway network from a given source to a destination \cite{shahzaad2021robust}. The composition of services creates a value-added service \cite{lakhdari2020Elastic,chaki2021adaptive,DBLP:journals/corr/abs-2107-12519}. We compose drone services to deliver packages while considering customer's QoS requirements. We assume that no handover of packages occurs among drones at intermediate stations as each drone has its own delivery plan, i.e., the same drone delivers a package from source to destination.

A key challenge in drone service composition is the uncertainty in congestion behaviour at recharging stations. This uncertainty is caused by the stochastic arrival of drones at particular stations. The arrival of a drone is greatly influenced by the payload weight, drone speed, and weather conditions \cite{DBLP:journals/corr/abs-2107-05173}. For example, several drones may be scheduled to arrive at a certain recharging station. If drones arrive earlier or later than the scheduled time, this may cause congestion at this station. A congested station is defined as a recharging station where all pads are occupied and the drone may have to wait for the availability of pads \cite{12}. Each drone that operates in a multi-drone environment has its own delivery plan. Therefore, an accurate prediction of congestion at stations may not be possible for long-term periods \cite{9284115}. This uncertainty in congestion behaviour makes the composition problem significantly complex compared to a static skyway network where all drone services are deterministic.

The existing drone service composition approaches do not consider the uncertainty in congestion behaviour at recharging stations \cite{alkouz2020swarm,8818436}. We propose a {\em top-k drone service composition framework} that can effectively deal with the uncertain nature of the environment. We assume that drones are partially recharged at intermediate recharging stations. This assumption helps drones delivering packages faster. First, we compute {\em top-k drone service compositions} based on the service time of each drone service without considering congestion conditions. Then, we rank the compositions based on the shortest service time to support the faster package delivery from a given source to a destination. For example, top-3 compositions are computed and ranked in ascending order based on the service time for each composition. We consider the probabilistic availability of pads to estimate the waiting and recharging times at each station under congestion conditions. Then, we compute a new delivery time for each composition in top-k compositions which is a sum of service time, waiting time, and recharging time for each drone service. We rerank the compositions using the new delivery times and select the best service composition plan. Finally, we compare the results of top-k composition approach with exhaustive drone service composition approach to analyze its performance in terms of execution time, delivery time, and delivery cost. We summarize the main contributions of this paper as follows:

\vspace{-0.2 cm}
\begin{itemize}
    \item[$\bullet$] Designing a system model for the provisioning of drone services.
    \item[$\bullet$] Proposing a top-k drone service composition framework under recharging constraints.
    \item[$\bullet$] Developing a heuristic-based approach for the composition of drone services and ranking the drone service composition.
    \item[$\bullet$] Conducting experiments using a real drone dataset to demonstrate the performance of the proposed composition approach.
\end{itemize}

\section{Motivating Scenario} \label{motivatingscenario}

\begin{figure}[t]

    \centering
    \includegraphics[width=0.7\textwidth]{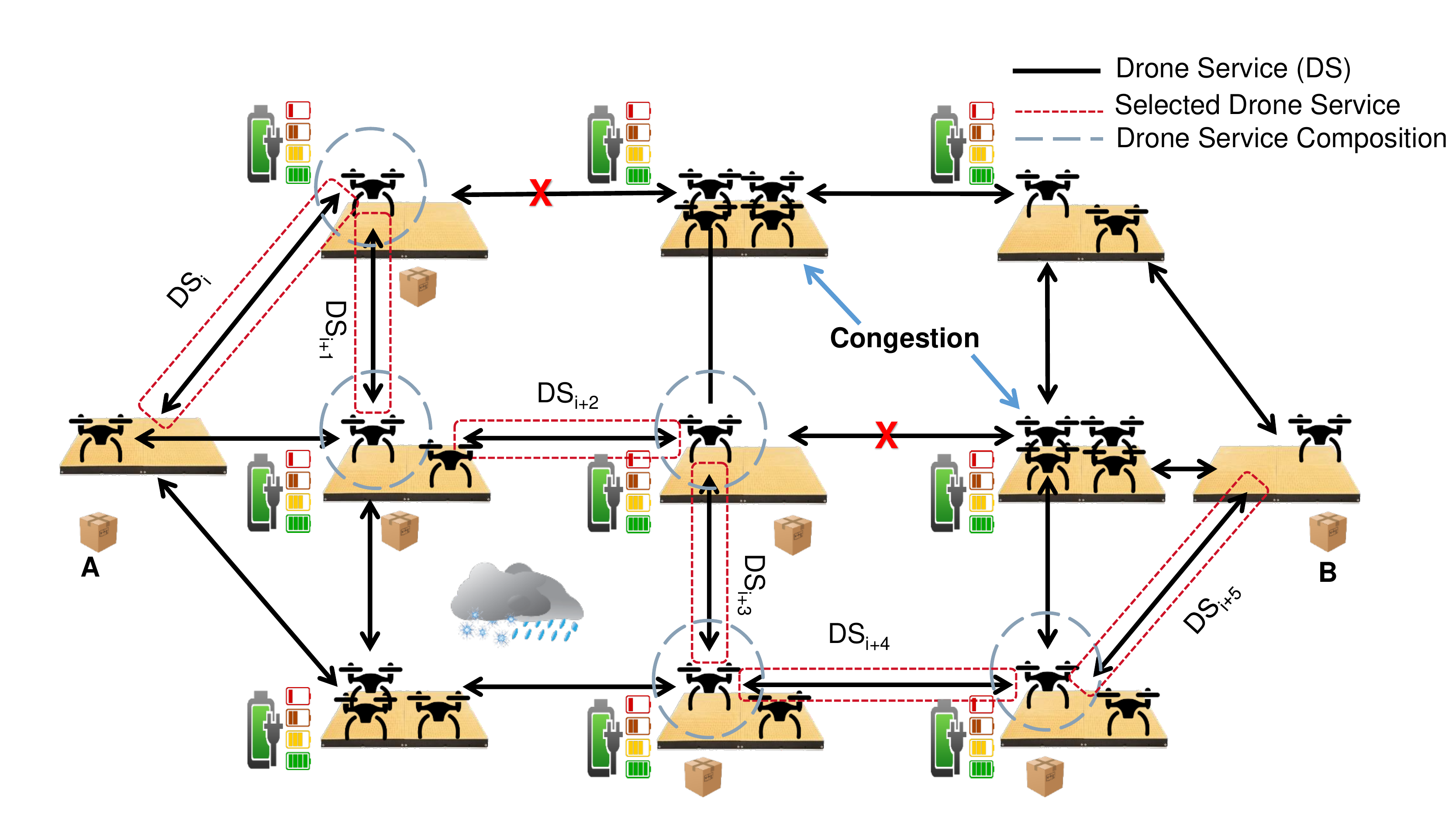}

\vspace{-0.4 cm}
    \caption{Drone service composition for package delivery in a skyway network}

    \label{fig1}

\vspace{-0.4 cm}
\end{figure}

Drones can be used for safe and contactless delivery of packages such as parcel, mail, medication, and meal. Let multiple delivery service providers offer package delivery services using drones in Texas, USA. Suppose Robert requests a fast package delivery service using a drone from \emph{San Marcos} to \emph{San Antonio} (92 km). The flight range of a typical delivery drone varies from 3 to 33 km \cite{12}. The weather conditions, the payload weight, and the drone speed influence the drone's flight range. Therefore, multiple times of recharge is required to meet the delivery request. In this regard, it is of paramount importance to avoid bad weather and congestion conditions at the recharging stations.

We construct a skyway network following the drone flying regulations such as avoiding flying in restricted areas (e.g., airport and military). The skyway network is divided into predefined skyway segments where each segment is a skyway path between two nodes. The nodes are the rooftops of the buildings which are assumed to be a \emph{recharging station} and/or a \emph{delivery target}. Each recharging station has a fixed set of pads for drones to land and recharge. Each skyway segment is a \textit{drone service} which is served by a drone. The nodes in the skyway network are considered as hubs where dynamic congestion of drones occurs, i.e., all recharging pads are occupied. An optimal drone service composition avoids hub nodes and provides fast and cost-efficient delivery. Fig. \ref{fig1} depicts a drone service composition scenario for package delivery from point A to point B.
\vspace{-0.4cm}
\section{Related Work} \label{relatedwork}

The existing research on drone-based deliveries can be divided into two categories: (1) Data-driven Approaches (2) Service-driven Approaches.

\textbf{Data-driven Approaches:} Data-driven approaches focus on point-to-point deliveries using drones \cite{7513397}. A parcel delivery system using drones is designed considering the impact of payload on the energy consumption of the drones \cite{torabbeigi2020drone}. The energy consumption of the drone is approximated as a linear function of the payload weight to schedule a reliable drone-based parcel delivery. It is assumed that the flying speed of drones is fixed. Strategic and operational planning is proposed for a given area based on a linear regression model. It is concluded that 60\% of flight paths fail to complete the delivery if the energy consumption is not considered. The proposed study focuses only on the flight time and payload weight as the factors affecting drone energy consumption. \textit{The proposed system does not consider the congestion conditions at recharging stations for drones}.

An energy consumption model is presented for automated drone delivery in \cite{choi2017optimization}. It is assumed that drones can perform multi-package deliveries in a predefined service area. The drone fleet size is optimized by analyzing the impact of payload weight and flight range considering battery capacity. They explore the relationship between four variables (working period, drone speed, demand density of service area, and battery capacity) to minimize the total costs of the drone delivery system. The study indicated that the long hours of operation would benefit both service providers and customers. They found that drone deliveries are more cost-effective in areas with high demand densities. \textit{This study does not take into account the recharging requirements of drones and the impact of congestion conditions on drone deliveries}.

A drone routing problem in a distribution network is studied considering the wind effects on power consumption \cite{Radzki2019}. It is assumed that the drone speed, wind speed, and wind direction remain constant during the delivery operation. It is also assumed that the payload weight of a drone remains fixed during a trip. As all influencing factors are deterministic, energy consumption becomes a constant number. All the deliveries are time-constrained, which requires the completion of the delivery operation in a given time window. Thus, energy constraints are ultimately transformed into generalized resource constraints. \textit{The proposed approach does not consider the real-world changing weather and congestion conditions at recharging stations in the drone delivery network.}

A modular optimization method is proposed for the drone delivery system in \cite{7934790}. The proposed method is beneficial in increasing the readiness of the drone fleet and decreasing the overall drone fleet size. A module in the proposed system lends more flexibility in drone operations with its interchangeable components: propellers, replaceable batteries, carriers, and motors.
A forward-looking strategy is applied to enhance the performance of drone-based delivery. The modular delivery drones are compared to non-modular delivery drones using the proposed approach. The simulation results demonstrate that the modular optimization method is more efficient at reducing the power consumption and delivery time of a drone. \textit{The proposed model does not consider the weather conditions and congestion conditions caused by other drones in the same delivery system}.

\textbf{Service-driven Approaches:} Service-driven approaches ensure \textit{congruent} and \textit{effective provisioning} of drone-based deliveries \cite{8818436,10.1007/978-3-030-33702-5_28,8355153,9099809}. There is a paucity of literature focusing on service-driven approaches that consider drone-based deliveries in complex and dynamic environments. A formal drone service model is designed considering the spatio-temporal features of the drone services in \cite{8818436}. The spatio-temporal features represent the location and time of the drone service. A formal QoS model is also designed to incorporate the non-functional properties of a drone service. The QoS properties include the service flight time and the delivery cost of a drone service. A heuristic-based algorithm is developed to select and compose the right drone services taking into account the QoS properties. \textit{The proposed approach focuses only on the deterministic properties of services which is not realistic.}

A prototype for drone service provision is presented that includes a drone, a controller, and a client \cite{8355153}. Each drone is embedded with a server to answer the service requests of clients made through a smartphone application. The simulation experiments are performed to analyze the factors that affect drone service delivery. The main factors include the number of drones, frequency of client requests, and relative localization of control stations. In addition, the scheduling strategies for distributing the service load among different drones are found more effective compared to a simple queue strategy. \textit{However, the different recharging requirements of drones and uncertain wind conditions are not considered.}

A deterministic drone service composition approach is proposed to incorporate the recharging constraints at stations in \cite{10.1007/978-3-030-33702-5_28}. The drone service selection and composition problem is formulated as a multi-armed bandit tree exploration problem. A skyline approach is proposed to reduce the search space for optimal selection of candidate drone services. A lookahead heuristic-based algorithm is presented for the selection and composition of optimal services. \textit{However, the uncertain weather conditions over different skyway segments and dynamic recharging constraints are not considered in the proposed composition approach.}

A drone service system is presented to provide long-distance delivery services considering refuelling and maintenance of drones in \cite{9099809}. The objective of this system is to minimize the travel distance of a drone and the number of landing depots during the delivery operation. An ant colony algorithm with the A* algorithm is proposed to solve the problem of long-distance delivery services. \textit{The proposed system does not take into account the factors affecting the flight range of a drone such as a payload and wind conditions.} Additionally, the congestion conditions at recharging stations are not considered in the proposed system. To the best of our knowledge, this paper is the first attempt to present a top-k drone service composition that considers congestion conditions at recharging stations.

\vspace{-0.4cm}
\section{Drone Service System Model} \label{problemdefinition}

We propose a drone service system model for the provisioning of drone delivery services. The proposed model consists of four components: (1) Skyway Network, (2) Drone Service Model, (3) No-Congestion Drone Service Model, and (4) Congestion Drone Service Model.
\vspace{-0.3 cm}
\subsection{Skyway Network}

We describe our multi-drone skyway network in which drone services operate to deliver packages. Let $D$ is a set of drones where $D = \{d_1, d_2,\ldots, d_n\}$. The skyway network is modelled as an undirected graph $G = (N, E)$. $N$ is a set of nodes, each of which represents a delivery target (i.e., customer's location) or recharging station. $E$ is a set of edges, each of which represents a skyway segment \textit{drone service} joining a pair of nodes. Each node has a \textit{fixed number of recharging pads}. $B$ is a set of battery capacities for all drones in $D$. The battery consumption and cost to travel from a node $i$ to $j$ are represented by $b_{ij}$ and $c_{ij}$ respectively. The battery consumption of the drone increases as the payload weight and the travelling distance increase.
\vspace{-0.3 cm}
\subsection{Drone Service Model}
The drone service, drone service query, and drone service composition problem are defined as follows.\\
\textbf{Definition 1: Drone Service (DS)}. A drone service is a tuple of $<DS\_id,$ $DS_{f}, DS_q>$, where
\begin{itemize}
    \item[$\bullet$] $DS.id$ is a unique drone service ID,
    \item[$\bullet$] $DS_{f}$ represents the delivery function of a drone over a skyway segment. The location and time of a drone service are tuples of $<loc_s,loc_e>$ and $<t_s, t_e>$, where
    \begin{itemize}
        \item $loc_s$ and $loc_e$ represent the start location and the end location of a drone service,
        \item $t_s$ and $t_e$ represent the start time and the end time of a drone service,
    \end{itemize}
    \item[$\bullet$] $DS_q$ is a tuple of $<q_1, q_2,\ldots,q_n>$, where each $q_i$ represents a quality parameter of a drone service, e.g., flight range and payload capacity.
\end{itemize}

\textbf{Definition 2: Drone Service Query (DSQ)}. A drone service query is defined as a service request for package delivery from a source location (i.e., warehouse rooftop) to a destination location (customer's building rooftop). A drone service query is a tuple $<\zeta, \xi, qt_{s}, w>$, where $\zeta$ is the source, $\xi$ is the destination, $qt_s$ is the query start time, and $w$ is the weight of the package.

\textbf{Definition 3: Drone Service Composition Problem}.
Given a set of drone services $S_{DS} = \{DS_1, DS_2,..., DS_n\}$ and drone service query $<\zeta, \xi, qt_s, w>$, the drone service composition problem is to compose the services for delivering a package from the source to the destination in minimum time.

\subsection{No-Congestion Drone Service Model}

\begin{figure}[t]

    \centering
    \includegraphics[width=0.9\textwidth, height=5cm]{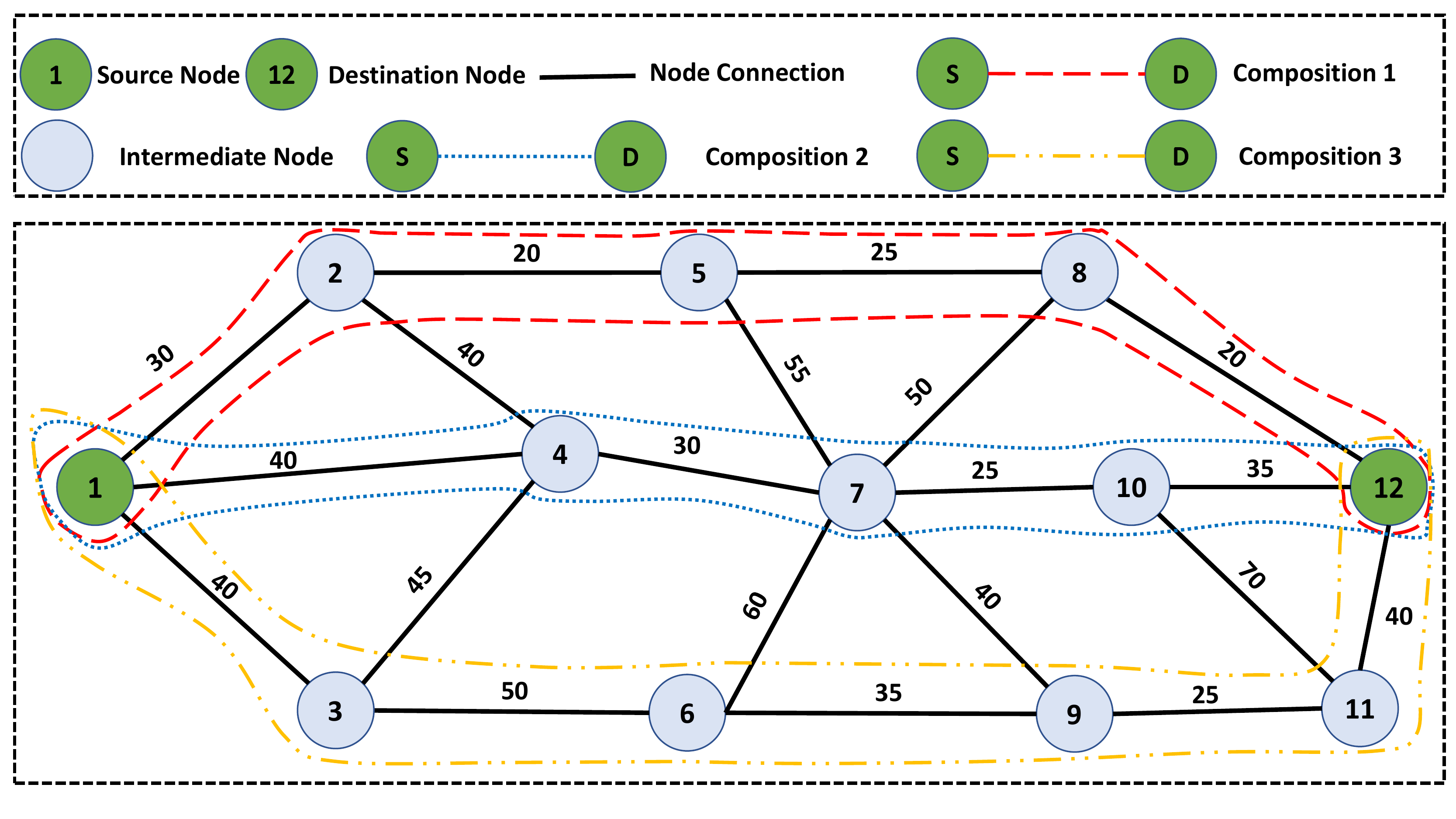}
    \vspace{-0.4 cm}

    \caption{Top-k compositions in no-congestion drone service model}
    
    \label{fig2}
    \vspace{-0.4 cm}
\end{figure}

We propose a no-congestion model for drone services where the congestion conditions at recharging stations are ignored. We assume that a recharging pad is \textit{always available} when a drone reaches a station, i.e., the recharging pads at each station are infinite. We use this assumption to compute \textit{top-k compositions} that consider only the service time of a drone service to select a service in the composition process. This is motivated by the fact that the service time is \textit{always higher} than the recharging time when a partial recharge policy is followed. The delivery time for a customer $C_n$ $(n \in N)$ is the sum of service times for each component drone service. Fig. \ref{fig2} represents top-K compositions in a no-congestion drone service model. The total delivery time for each composition is calculated by adding the sum of deterministic service times for all component drone services. For example, the service time for a drone service from node 1 to node 2 is 30 minutes. The total delivery time is the sum of services leading from source node 1 to destination node 12 which is 95 minutes in case of \textit{service composition 1}.

\subsection{Congestion Drone Service Model}

Congestion is a natural phenomenon in a resource-constrained dynamic network \cite{8737723}. The effect of congestion is primarily that the waiting time on a congested recharging station increases as more drones approach the same congested station. In a deterministic skyway network, each drone service has perfect knowledge of all other incoming and outgoing drones at a particular station and their scheduled arrival times. Each drone service then chooses the recharging station with the lowest waiting time. In this regard, the delivery time for a customer $C_n$ $(n \in N)$ can simply be modelled as follows:

\vspace{-0.4 cm}
\begin{equation}\label{eq:2}
  T_n = S_n + R_n + W_n
\end{equation}

where $T_n$ is the deterministic delivery time, $S_n$ is the service time of all component drone services in the skyway path from source to destination, and $R_n$ and $W_n$ are the sums of recharging and waiting times at each intermediate station, respectively.

In a dynamic skyway network, the drones do not have perfect knowledge about the availability of pads at a recharging station. Therefore, we consider the likelihood of pad's availability at a recharging station, i.e., probability of availability. We compute the delivery time considering the probabilities of the pad's availability and its duration of availability for recharging. We use the following equation to calculate the delivery time for a customer $C_n$ $(n \in N)$:

\vspace{-0.4 cm}
\begin{equation}\label{eq:3}
  T_n = S_n + \sum_{i=0}^n Pr_i * (R_i + W_i)
\end{equation}

where $T_n$ is the stochastic delivery time, $S_n$ is the service time of all component drone services in the skyway path from source to destination, $R_i$ and $W_i$ are the recharging and waiting times at a station $i$, and $Pr_i$ is the probability of recharging and waiting times for a station $i$.

\begin{figure}[t]

    \centering
    \includegraphics[width=0.9\textwidth, height=5cm]{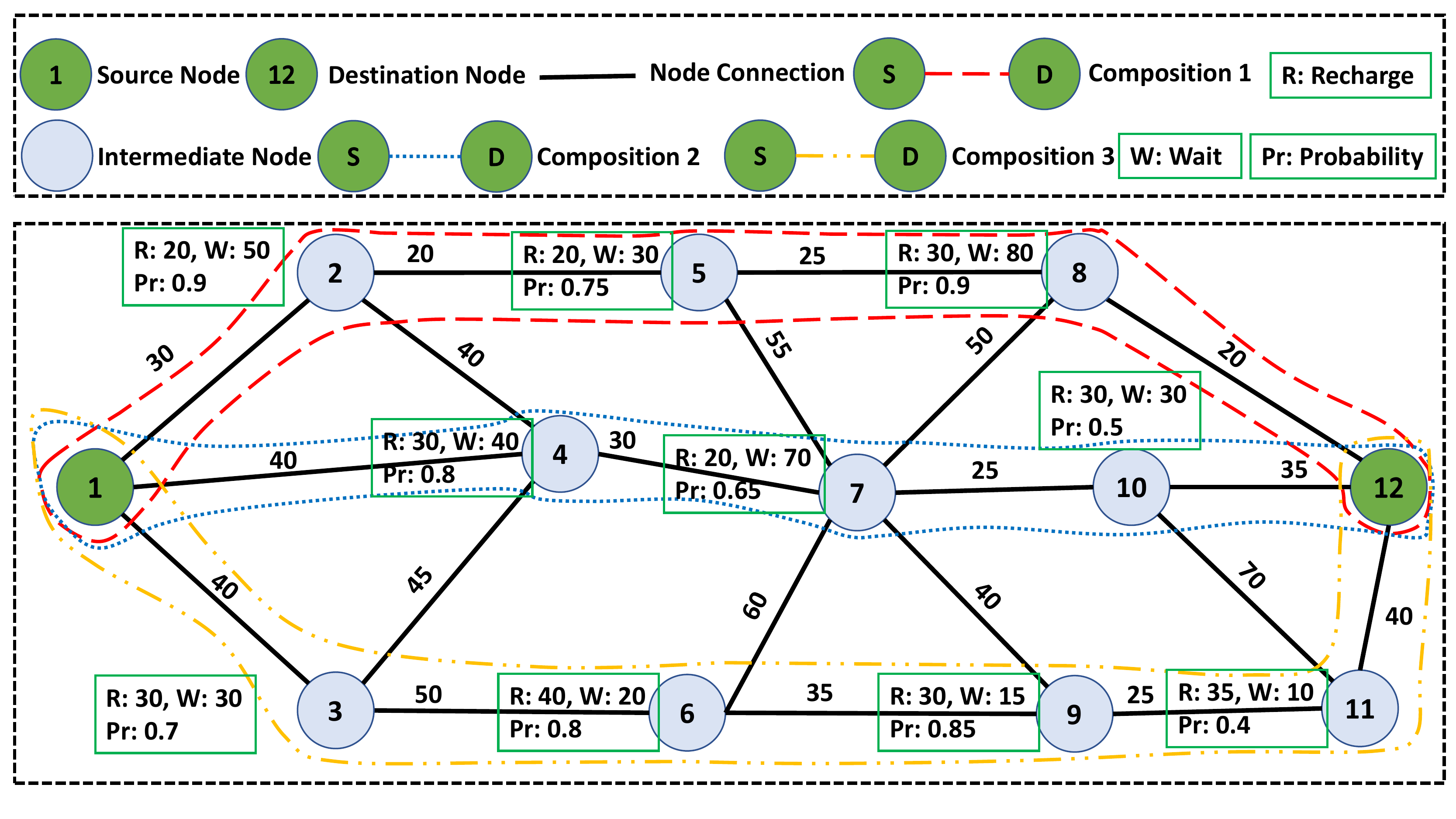}
    \vspace{-0.4 cm}

    \caption{Extended top-k composition in congestion drone service model}

    \label{fig3}
    \vspace{-0.4 cm}

\end{figure}

The probabilistic availability of recharging stations varies with time. Therefore, we compute probabilities incrementally for neighbour recharging stations corresponding to the current station during the delivery operation. This process continues until the package is delivered to its desired destination. We incorporate the recharging and waiting times with their probabilities in precomputed top-k compositions as shown in Fig. \ref{fig3}.

\vspace{-0.3 cm}
\section{Top-k Drone Service Composition Framework} \label{selection}

A single drone service usually cannot satisfy a user’s end-to-end delivery requirements. We often need to use drone service composition, which aggregates a set of drone services in order to serve a long-distance delivery request. In this paper, we propose a top-k drone service composition approach to support long-distance package deliveries using drones. We initially compute and rank top-k compositions considering service times of each component drone service. The service time represents the time to travel from one end to another end of a drone service.

In real-world situations, drone service compositions are influenced by the stochastic arrival of drones at intermediate stations and the changes in weather conditions. As a result, the established composition plans may become non-optimal. We, therefore, consider uncertainties involved in weather and congestion conditions at stations to provide an efficient and reliable drone service composition. We use the probabilistic arrival of drones at recharging stations that are a part of the top-k compositions. In addition, we also compute their effects on waiting and recharging times. As the congestion conditions are time-variant, we compute probabilities at each recharging station incrementally. We then calculate the stochastic delivery time using equation \ref{eq:3} in the congestion drone service model to incorporate the effects of dynamic congestion conditions in top-k compositions. We rerank the extended composition plans based on the delivery times with higher probabilities.
\vspace{-0.3 cm}
\subsection{Algorithm}
This section describes the top-k drone service composition algorithm for drone-based delivery services. The drone service composition process is accomplished in two phases. In the first phase, we compute, select, and rank top-k compositions considering the delivery time in no-congestion drone service model. In the second phase, we incorporate dynamism in top-k compositions considering congestion conditions described in congestion drone service model and rerank the extended composition plans. The details of the algorithm are described in Algorithm \ref{alg:algorithm1}.

\begin{algorithm}[t]
 \small
 \caption{Top-k Drone Service Composition}\label{alg:algorithm1}
	\begin{algorithmic}[1]
    \REQUIRE
		$G$, $D$, $\zeta$, $\xi$, $w$, $qt_{s}$, $k$
	\ENSURE
		$DSComp$\\
		\STATE $DSComp \gets \phi$
		\STATE $DSBase \gets \phi$
		\STATE $topKComp \gets \phi$
		\STATE $T_{DSComp} \gets \phi$
		\STATE $T_{DSBase} \gets \phi$
		\STATE $d_{sel} \gets$ block\_nested\_loop ($D$, $w$)
        \item[]
        \textbf{Phase 1. Initial top-k compositions}
		\STATE $topKComp, T_{DSBase} \gets$ base\_comps ($G$, $\zeta$, $\xi$, $k$, $d_{sel}$, $w$)
		\STATE $DSBase \gets$ rank\_comps ($topKComp$, $T_{DSBase}$)
        \item[]
		\textbf{Phase 2. Extend top-k compositions considering congestion conditions}
		\FOR{$DS_i \in DSBase$}
		\STATE $curTime \gets  qt_{s}$
		\STATE $T_i \gets 0$
		\FOR{$ds_j \in DS_i$}
		\STATE $Pr_j, W_j, R_j \gets $ probability\_wait\_recharge ($ds_j.loc_e$, $curTime$)
		\STATE $T_i \gets T_i + S_j + Pr_j * (R_j + W_j)$
		\STATE $curTime \gets  curTime + T_i$
		\ENDFOR
		\STATE $T_{DSComp}$.append($T_i$)
		\ENDFOR
		\STATE $DSComp \gets$ rank\_comps ($DSBase$, $T_{DSComp}$)
		\RETURN $DSComp$
	\end{algorithmic}
\end{algorithm}
\setlength{\textfloatsep}{4pt}
\vspace{-0.8 cm}
In Algorithm \ref{alg:algorithm1}, the output $DSComp$ is a set of top-k drone service composition plans from a source location to a destination location. The input is the skyway network represented by graph $G$, the set of delivery drones $D$, the source $\zeta$, the destination $\xi$, the package weight $w$, the query start time $qt_{s}$, and the number of top compositions to be selected $k$. Each skyway segment drone service in graph $G$ is served by a drone selected from the drone set $D$. We consider the start location, end location, and the distance between the two ends of each skyway segment drone service in graph $G$.
We create empty lists for $DSComp$, $DSBase$, $topKComp$, $T_{DSComp}$, and $T_{DSBase}$ (Lines 1-5). We use the \emph{Block Nested Loop (BNL)} \cite{6916872} algorithm to select an optimal set of drones from a large set of delivery drones $D$ given the payload weight (Line 6). Algorithm \ref{alg:algorithm2} provides the details of the BNL algorithm. Multiple drone service providers offer package delivery services. Each provider has several drones with different quality attributes. The BNL approach supports the selection of an optimal drone set determined to be a good fit for the delivery request. First, we filter the large set of delivery drones $D$ based on the package weight $w$ to select the candidate drones in Algorithm \ref{alg:algorithm2} (Lines 2-6). Then, we select a set of non-dominated drones based on the best QoS properties for each candidate drone. We use the negative and positive parameters to select drones, such as recharging time and travel distance, respectively (Lines 8-9). We obtain the better and worse values of quality parameters for each drone using Algorithm \ref{alg:algorithm3}. The range of a drone is of paramount importance to serve long-distance areas. Therefore, we prefer the flight range parameter for selecting a drone from the optimal drone set to serve the delivery request in Algorithm \ref{alg:algorithm2} (Line 31).

\begin{algorithm}[t]
 \small
 \caption{block\_nested\_loop ($D$, $w$)}\label{alg:algorithm2}
    \begin{algorithmic}[1]
    \STATE $candidateDrone \gets \phi$
    \FOR{each $drone \in D$}
    \IF{$drone.pl \geq w$}
    \STATE $candidateDrone$.append($drone$)
    \ENDIF
    \ENDFOR
    \STATE $rows \gets candidateDrone.to\_dict()$
    \STATE $to\_min \gets $ negative parameters, e.g., recharging time
    \STATE $to\_max \gets $ positive parameters, e.g., travel distance
    \STATE $to\_sel \gets$ important parameter for drone selection, e.g., range
    \STATE $selDrone \gets candidateDrone[0]$
    \FOR{each $drone \in candidateDrone[1:n]$}
    \STATE $is\_dominated \gets False$
    \STATE $to\_drop \gets$ set()
    \FOR{each $q_i \in selDrone$}
    \STATE $better, worse \gets$ count\_diff ($rows$[$drone.q_i$], $rows$[$q_i$], $to\_min$, $to\_max$)
    \IF{$worse > 0$ and $better = 0$}
    \STATE $is\_dominated \gets True$
    \STATE break
    \ENDIF
    \IF{$better > 0$ and $worse = 0$}
    \STATE $to\_drop$.add($q_i$)
    \ENDIF
    \IF{$is\_dominated$}
    \STATE continue
    \ENDIF
    \STATE $selDrone \gets selDrone$.difference($to\_drop$)
    \STATE $selDrone$.add($drone$)
    \ENDFOR
    \ENDFOR
    \RETURN $selDrone [to\_sel]$
    \end{algorithmic}
\end{algorithm}

In phase 1, we compute top-k compositions and their delivery times using the selected drone and payload weight. Each drone service composition constitutes a skyway path based on the shortest delivery time leading the package $w$ from the source $\zeta$ to the destination $\xi$ (Line 7). We perform a straightforward ranking of compositions considering the respective delivery times (Line 8). In phase 2, we calculate the wait and recharge times and their corresponding probabilities at certain timestamps using a black-box approach for each component service and its intermediate station. We repeat the process for all top-k compositions and estimate the updated times considering weather and congestion conditions (Lines 9-18). We rerank the composition plans based on their extended delivery times and finally return a list of top-k drone service compositions (Lines 19-20). The reranking is essential as an initial optimal composition plan may become non-optimal due to changing weather and congestion conditions.

\begin{algorithm}[t]
 \small
 \caption{count\_diff ($paramA$, $paramB$, $to\_min$, $to\_max$)}\label{alg:algorithm3}
    \begin{algorithmic}[1]
    \STATE $better \gets 0$, $worse \gets 0$
    \FOR{each $f \in to\_min$}
    \STATE $better \gets better + (paramA[f] < paramB[f]$)
    \STATE $worse \gets worse + (paramA[f] > paramB[f]$)
    \ENDFOR
    \FOR{each $f \in to\_max$}
    \STATE $better \gets better + (paramA[f] > paramB[f]$)
    \STATE $worse \gets worse + (paramA[f] < paramB[f]$)
    \ENDFOR
    \RETURN $better, worse$
    \end{algorithmic}
\end{algorithm}
\vspace{-0.4 cm}
\section{Performance Evaluation}
\vspace{-0.2 cm}
We evaluate the performance of our proposed drone service composition approach using the following evaluation settings:
\begin{itemize}
    \item[$\bullet$] \textbf{Performance Metrics:} The delivery time and cost are paramount in drone delivery services. We use the drone travelling distance as a function of delivery cost. Therefore, we use (1) \textit{execution time}, (2) \textit{delivery time}, and (3) \textit{distance travelled} as performance metrics. The execution time is used to evaluate the runtime complexity of the algorithms.
    \item[$\bullet$] \textbf{Baseline:} To evaluate our proposed approach, we compare the top-k drone service composition algorithm with an exhaustive drone service composition approach. The exhaustive composition approach takes exponential time for the increasing number of nodes.
\end{itemize}

\vspace{-0.6 cm}
\subsection{Experiment Settings with Real-world Datasets}

\begin{figure}[t]

    \centering
    \includegraphics[width=0.9\textwidth]{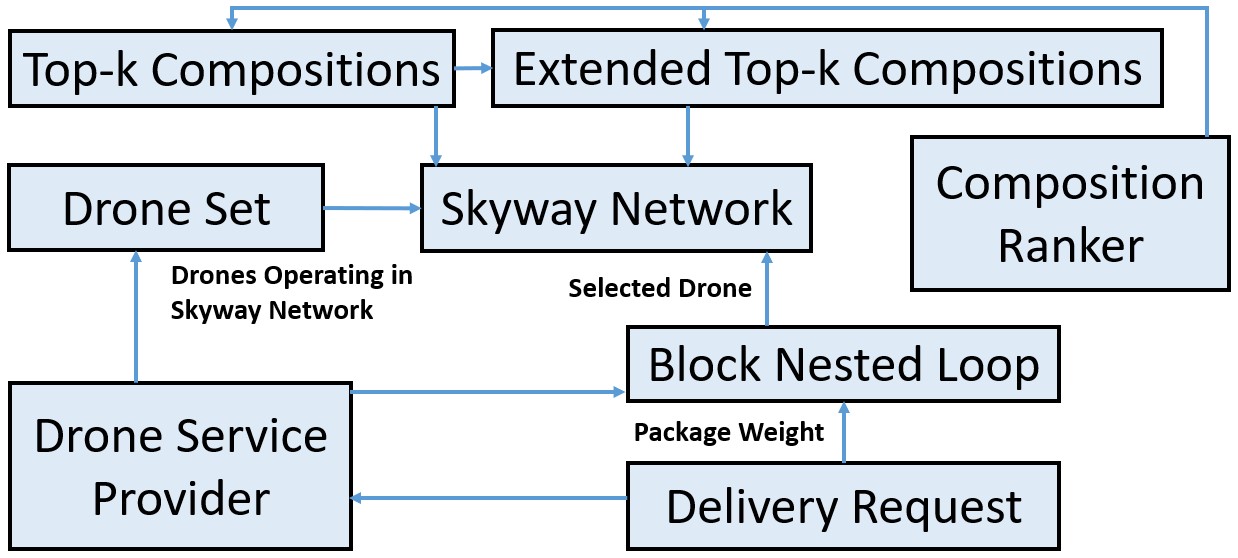}
    \vspace{-0.2 cm}

    \caption{Top-k drone service composition framework for delivery services}

    \label{fig:fig4.jpg}
\end{figure}

We develop a top-k drone service composition framework for delivery services to evaluate the performance of our proposed approach. The modules of the framework are shown in Fig.~\ref{fig:fig4.jpg}. We build a skyway network using the NetworkX python library, where each node can be a delivery target or a recharging station. We model multiple drone services from different drone service providers operating in the same network. The drone set consists of quality parameters of each drone operating in the skyway network, e.g., flight range and payload capacity. The experiments are conducted for an average of 50\% times the total number of nodes. For example, if there are 40 nodes in the network, the experiment is performed 20 times. We select a random source and a random destination point for each experiment. The delivery request module is used for initiating drone-based delivery services. The Block Nested Loop implements Algorithm \ref{alg:algorithm2} which is used to select the right drone.
We use a real urban road network dataset for the Tokyo city, including data for coordinates, nodes, and length of each edge between two nodes \cite{karduni2016protocol}. We extract a sub-network of 5000 connected nodes to construct a skyway network. We augment a dataset for different types of drones considering the payload, speed, flight range, recharging time, and battery capacity. The experimental variables are described in Table \ref{tab:table1}.

\begin{table}
\centering
\caption{Experimental Variables}
\label{tab:table1}
\begin{tabular}{|l|l|}
\hline
 \textbf{Variable} &  \textbf{Values} \\

\hline

Drone model &  DJI M200 V2\\ \hline

Maximum payload capacity & 1.45 Kg \\ \hline

Maximum drone flight time & 24 min \\ \hline

Maximum drone flight range & 32.4 km \\ \hline

Maximum drone speed & 81 km/h \\ \hline

Recharging time from 0\% to 100\% & 2.24 hours \\ \hline

Maximum nodes in the skyway network &  40  \\ \hline

No. of pads at each recharging station & 3 \\ \hline

Experiment run the total number of nodes & 50\% \\

\hline
\end{tabular}
\end{table}

\subsection{Results and Discussion}

The proposed top-k approach performs the composition of the right drone services to deliver the package faster. We rank the top-k compositions and select the best composition plan for comparison with the baseline approach. For example, the top-3 compositions in the results show the best extended composition plan among 3 compositions with the least delivery time that is computed after incorporating the probabilistic recharging and waiting times. A similar approach is considered for top-4 and top-5 compositions that constitute initial 4 and 5 skyway paths from a given source to a destination with minimum delivery time.

\vspace{-0.2 cm}
\subsubsection{Average Execution Time.}
\begin{figure} [t]
    \centering
    \begin{minipage}{0.5\textwidth}
        \includegraphics[width=\textwidth]{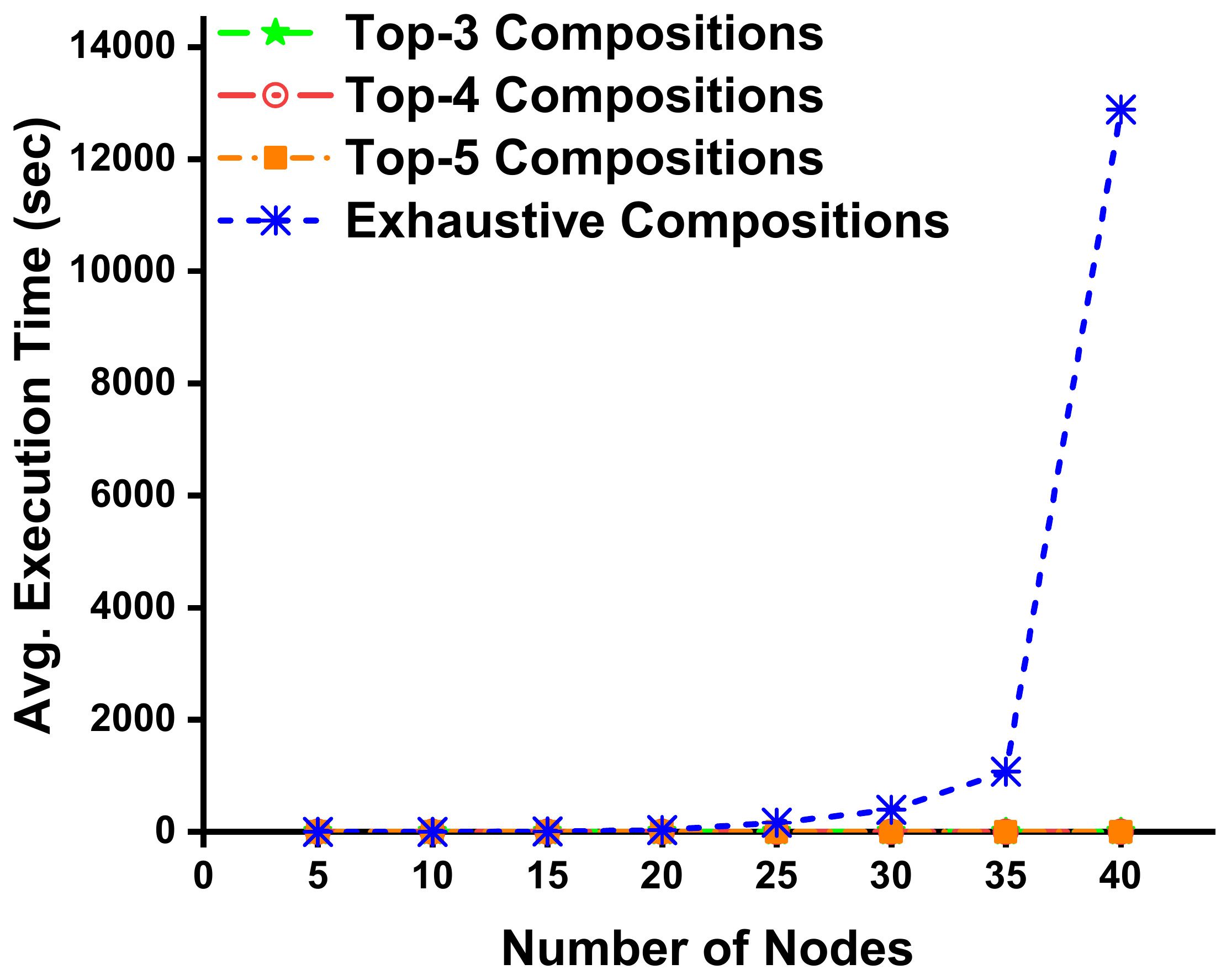} 
        \vspace{-0.4cm}
        \caption{Average execution time}
        \label{fig:fig5.pdf}
    \end{minipage}\hfill
    \begin{minipage}{0.5\textwidth}
        \includegraphics[width=\textwidth]{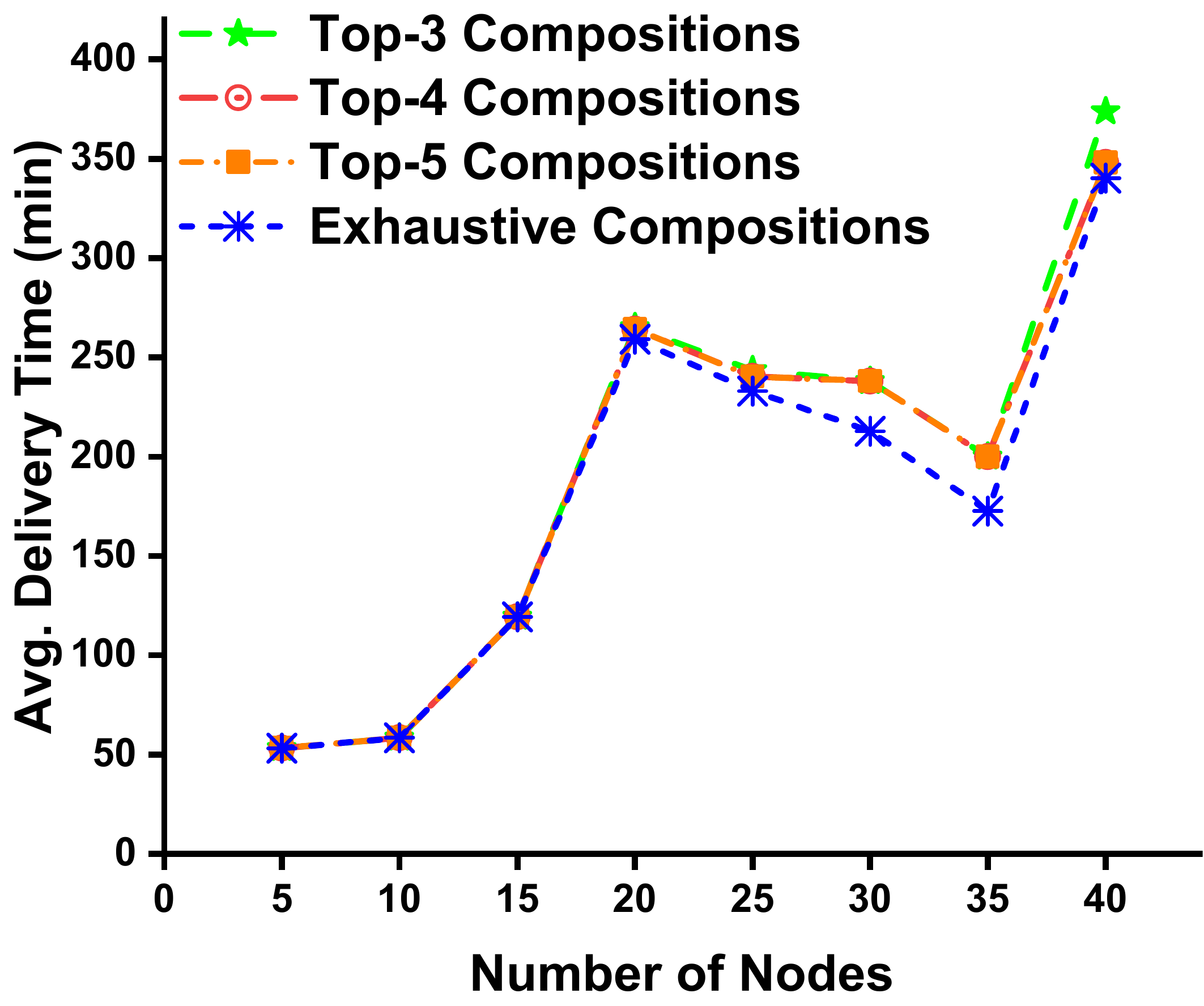} 
        \vspace{-0.4cm}
        \caption{Average delivery time}
        \label{fig:fig6.pdf}
    \end{minipage}
\end{figure}

The time complexity is an important parameter to evaluate the performance of an algorithm. The exhaustive composition approach is computationally expensive compared to the proposed top-k composition approach. The execution time increases as the number of possible drone service compositions increase. The average execution times for exhaustive, top-3, top-4, and top-5 compositions are presented in  Fig.~\ref{fig:fig5.pdf}. The execution times for all top-k compositions are approximately similar because of avoiding exhaustive drone service compositions. As expected, the average execution time for the exhaustive grows exponentially for an increasing number of nodes. The experiments indicate that when the nodes are above 40, the results' trends are similar. As a result, we set the maximum number of nodes at 40. It shows that the use of the baseline approach is not practical in real-world scenarios for large-scale problems because of its exhaustive nature. We observe that our proposed approach outperforms the exhaustive composition approach to compute an optimal composition plan.

\vspace{-0.4cm}

\subsubsection{Average Delivery Time.}
The delivery time of a drone is a summation of recharging, waiting, and service times. The delivery time is mainly affected by the occupancy of certain recharging stations for long periods of time. Fig.~\ref{fig:fig6.pdf} shows the delivery times of exhaustive, top-3, top-4, and top-5 compositions. The exhaustive approach always computes all possible drone service compositions, which in turn provides exact solutions. The top-k compositions provide delivery solutions close to the exhaustive composition approach. We observe that the delivery time is 5\% higher for the top-3 compositions and 4\% higher for the top-4 and top-5 compositions compared to the exhaustive composition approach. This increase in delivery time is because the top-k compositions do not initially anticipate the arrival of other drones and congestion conditions at recharging stations. However, the top-k composition approach is significantly faster than the exhaustive composition approach, as shown in  Fig.~\ref{fig:fig5.pdf}.

\vspace{-0.4 cm}
\subsubsection{Average Distance Travelled.}
\begin{figure}[t]

    \centering
    \includegraphics[width=0.5\textwidth]{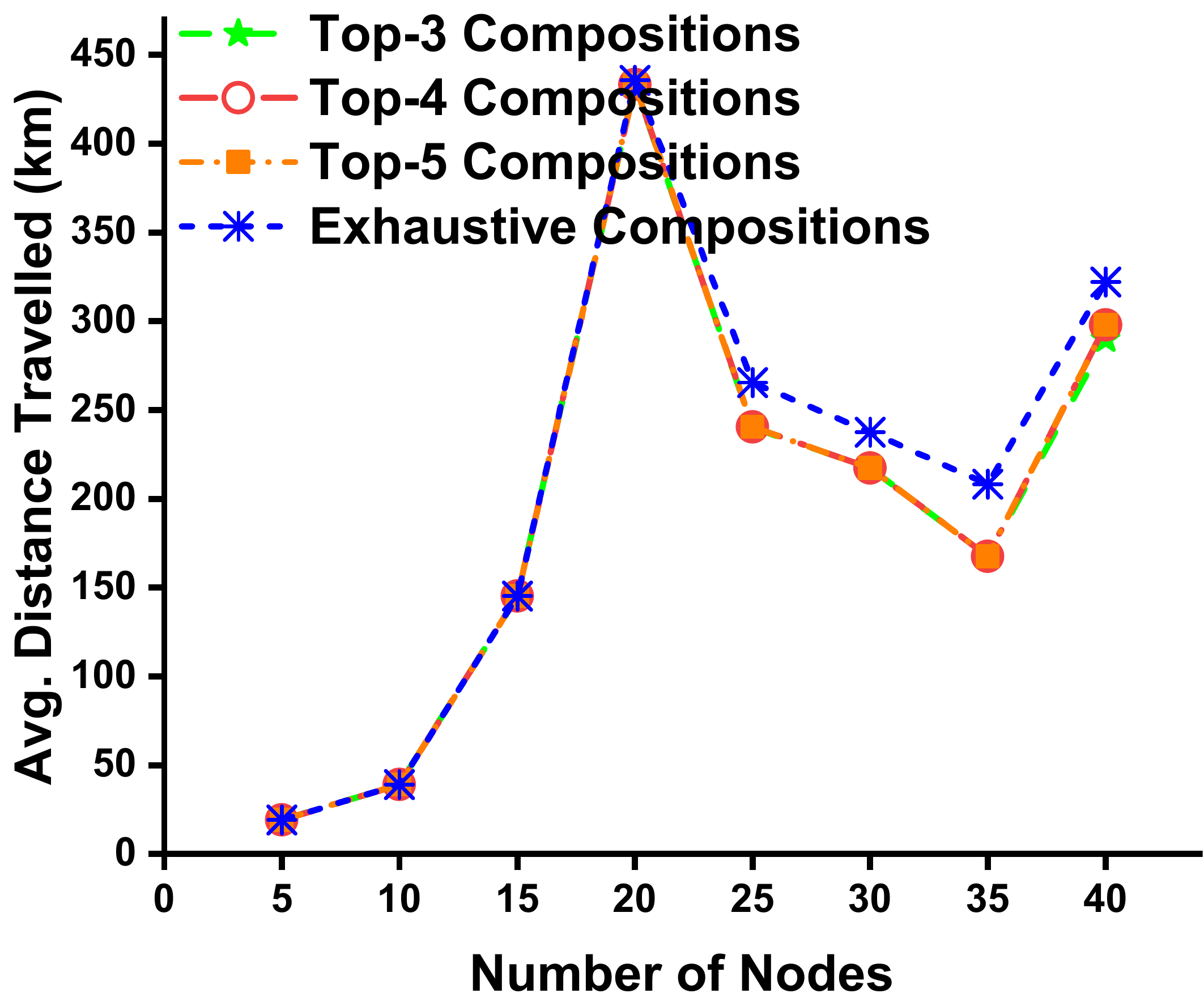}
    \vspace{-0.4 cm}
    \caption{Average distance travelled}
    \label{fig:fig7.pdf}
\end{figure}

The cost of drone-based delivery services is estimated to be \$0.1 for a 2 kg package delivery within a range of 10 km \cite{5}. We define the cost function of the drone delivery as its travelling distance. Due to the uncertain nature of the environment, the initially attractive services may lead to congested stations. The average distances travelled by exhaustive, top-3, top-4, and top-5 compositions are shown in Fig.~\ref{fig:fig7.pdf}. The least distance services selected by the top-k composition approach may result in higher delivery time because of the uncertainty involved in the composition process. We observe that the distance travelled by the exhaustive composition approach is slightly higher than our proposed top-k composition approach. This is because the exhaustive composition approach always selects the optimal delivery time services. It shows that the delivery cost for the top-k composition approach is slightly less than the exhaustive composition approach.

\section{Conclusion}

We propose a novel framework for drone service composition considering the stochastic congestion constraints at recharging stations. A Block Nested Loop algorithm is used for the selection of the right drone at the source location. The proposed approach initially computes top-k compositions with minimum service times from the source to the destination. Then, we incorporate the probabilistic impact of recharging time and waiting time at stations. We rank the top-k compositions based on their delivery times and select the best composition plan. We run a set of experiments to evaluate the efficiency of our approach compared to the exhaustive composition approach. The experimental results prove that the proposed approach is computationally efficient and cost-effective to deliver the packages compared to the exhaustive composition approach. Moreover, our proposed approach is a practical solution for real-world scenarios of drone delivery services due to its stable and computationally efficient solutions. In future, we plan to include other types of environmental uncertainties such as temperature and their impact on drone delivery.

\vspace{-0.2 cm}
\subsubsection*{Acknowledgment.}
This research was partly made possible by DP160103595 and LE180100158 grants from the Australian Research Council. The statements made herein are solely the responsibility of the authors.

%
%
%

\bibliographystyle{splncs04}
\bibliography{references}

\begin{thebibliography}{10}
\providecommand{\url}[1]{\texttt{#1}}
\providecommand{\urlprefix}{URL }
\providecommand{\doi}[1]{https://doi.org/#1}

\bibitem{8682048}
Shakhatreh, H., et~al.: Unmanned aerial vehicles (uavs): A survey on civil
  applications and key research challenges. IEEE Access  \textbf{7},
  48572--48634 (2019)

\bibitem{2021335}
Shahzaad, B., et~al.: Resilient composition of drone services for delivery.
  Future Gener. Comput. Syst.  \textbf{115},  335--350 (2021)

\bibitem{9086010}
{Chamola}, V., et~al.: A comprehensive review of the covid-19 pandemic and the
  role of iot, drones, ai, blockchain, and 5g in managing its impact. IEEE
  Access  \textbf{8},  90225--90265 (2020)

\bibitem{Aurambout2019}
Aurambout, J.P., Gkoumas, K., Ciuffo, B.: Last mile delivery by drones: an
  estimation of viable market potential and access to citizens across european
  cities. Eur. Transp. Res. Rev.  \textbf{11} (2019)

\bibitem{Bouguettaya:2017:SCM:3069398.2983528}
Bouguettaya, A., et~al.: A service computing manifesto: The next 10 years.
  Commun. ACM  \textbf{60}(4),  64--72 (2017)

\bibitem{alkouz2020formation}
Alkouz, B., Bouguettaya, A.: Formation-based selection of drone swarm services.
  In: Mobiquitous. EAI (2020)

\bibitem{10.1145/3460418.3479289}
Lee, W., et~al.: Package delivery using autonomous drones in skyways. In: Proc.
  UbiComp/ISWC. p. 48–50 (2021)

\bibitem{shahzaad2021robust}
Shahzaad, B., Bouguettaya, A., Mistry, S.: Robust composition of drone delivery
  services under uncertainty. In: IEEE ICWS (2021)

\bibitem{lakhdari2020Elastic}
Lakhdari, A., et~al.: Elastic composition of crowdsourced iot energy services.
  In: Mobiquitous. EAI (2020)

\bibitem{chaki2021adaptive}
Chaki, D., Bouguettaya, A.: Adaptive priority-based conflict resolution of iot
  services. arXiv preprint arXiv:2107.08348  (2021)

\bibitem{DBLP:journals/corr/abs-2107-12519}
Lakhdari, A., Bouguettaya, A.: Proactive composition of mobile iot energy
  services. In: IEEE ICWS (2021)

\bibitem{DBLP:journals/corr/abs-2107-05173}
Alkouz, B., Bouguettaya, A.: Provider-centric allocation of drone swarm
  services. In: IEEE ICWS (2021)

\bibitem{12}
Kim, J., et~al.: Cbdn: Cloud-based drone navigation for efficient battery
  charging in drone networks. Trans. Intell. Transp. Syst. pp. 1--18 (2018)

\bibitem{9284115}
Shahzaad, B., Bouguettaya, A., Mistry, S.: A game-theoretic drone-as-a-service
  composition for delivery. In: IEEE ICWS. pp. 449--453 (2020)

\bibitem{alkouz2020swarm}
Alkouz, B., Bouguettaya, A., Mistry, S.: Swarm-based drone-as-a-service (sdaas)
  for delivery. In: IEEE ICWS. pp. 441--448 (2020)

\bibitem{8818436}
Shahzaad, B., et~al.: Composing drone-as-a-service (daas) for delivery. In:
  IEEE ICWS. pp. 28--32. Milan, Italy (2019)

\bibitem{7513397}
{Dorling}, K., et~al.: Vehicle routing problems for drone delivery. Trans.
  Syst., Man, Cybern.  \textbf{47}(1),  70--85 (2017)

\bibitem{torabbeigi2020drone}
Torabbeigi, M., Lim, G.J., Kim, S.J.: Drone delivery scheduling optimization
  considering payload-induced battery consumption rates. J. Intell. Robot.
  Syst.  \textbf{97}(3),  471--487 (2020)

\bibitem{choi2017optimization}
Choi, Y., Schonfeld, P.M.: Optimization of multi-package drone deliveries
  considering battery capacity. In: 96th Annual Meeting of the Transportation
  Research Board. pp. 8--12. Washington, DC, USA (2017)

\bibitem{Radzki2019}
Radzki, G., Thibbotuwawa, A., Bocewicz, G.: Uavs flight routes optimization in
  changing weather conditions - constraint programming approach. Appl. Comput.
  Sci.  \textbf{15}(3),  5--20 (2019)

\bibitem{7934790}
{Lee}, J.: Optimization of a modular drone delivery system. In: IEEE SysCon.
  pp.~1--8 (2017)

\bibitem{10.1007/978-3-030-33702-5_28}
Shahzaad, B., et~al.: Constraint-aware drone-as-a-service composition. In:
  ICSOC. pp. 369--382. Springer (2019)

\bibitem{8355153}
Alwateer, M., Loke, S.W., Rahayu, W.: Drone services: An investigation via
  prototyping and simulation. In: IEEE WF-IoT. pp. 367--370 (2018)

\bibitem{9099809}
Shao, J., et~al.: A novel service system for long-distance drone delivery using
  the “ant colony+a*” algorithm. IEEE Syst. J. pp. 1--12 (2020)

\bibitem{8737723}
{Fotouhi}, Z., et~al.: A general model for ev drivers’ charging behavior.
  IEEE Trans. Veh. Technol.  \textbf{68}(8),  7368--7382 (2019)

\bibitem{6916872}
Hsu, W.T., et~al.: Skyline travel routes: Exploring skyline for trip planning.
  In: IEEE MDM. vol.~2, pp. 31--36. QLD, Australia (2014)

\bibitem{karduni2016protocol}
Karduni, A., Kermanshah, A., Derrible, S.: A protocol to convert spatial
  polyline data to network formats and applications to world urban road
  networks. Sci. Data  \textbf{3}(1),  160046 (2016)

\bibitem{5}
D'Andrea, R.: Guest editorial can drones deliver? Trans. Autom. Sci. Eng.
  \textbf{11}(3),  647--648 (2014)

\end{thebibliography}
\vspace{-0.4 cm}
\end{document}